\documentstyle[bo99,epsfig]{article}

\title{Magnetic Field Limit on SGR 1900+14}

\author{R.E. Rothschild $^1$, D. Marsden$^{1,2}$, \& R.E. Lingenfelter $^1$}

\affil{1) Center for Astrophysics and Space Sciences, University of California, San Diego, La Jolla, CA USA,\\
2) presently NAS/NRC Research Associate at Goddard Space Flight Center}

\begin{document}

\maketitle

\begin{abstract}

We measured the period and spin-down rate for SGR 1900+14 during the
quiescient period two years before the recent interval of renewed burst
activity.  We have shown that the spin-down age of SGR 1900+14 is
consistent with a braking index of $\sim$1 which is appropriate for
wind torques and not magnetic dipole radiation.  We have shown that a
combination of dipole radiation, and wind luminosity, coupled with
estimated ages and present spin parameters, imply that the magnetic
field for SGR 1900+14 is less than 6$\times$10$^{13}$ G and that the
efficiency for conversion of wind luminosity to x-ray luminosity is
$<$2\%.

\keywords{gamma-rays: bursts, pulsars: individual (SGR 1900+14), 
magnetic fields}

\end{abstract}

\section{ Spin-Down History of SGR 1900+14}

The spin-down of SGR 1900+14 from 1966 September to 1999 April (Figure
1) is characterized by three intervals of time for which the
spin-down rate was essentially constant within the interval.
This characterization of the SGR 1900+14 spin-down is based upon either
direct measurements of $\dot{P}$ as part of the period determination,
or upon differences in measured spin periods between two different
observations.  The first interval begins with the RXTE observation in
September of 1996 and ends with the ASCA observation at the beginning
of May, 1998.  The mean spin-down $\dot{P} \sim$ 6$\times$10$^{-11}$
s/s (Marsden, Rothschild and Lingenfelter 1999a; Hurley et al. 1999;
Woods et al. 1999a).  The second interval begins with the onset of
bursting on May 26, 1998 and continues until mid-September 1998.  The
mean spin-down during this time $\dot{P} \sim$13$\times$10$^{-11}$ s/s
(Kouveliotou et al. 1999; Marsden, Rothschild and Lingenfelter 1999a;
Murakami et al. 1999).  The third interval begins in mid-September 1998
and continues at least until March 30, 1999.  The mean spin-down at
that time $\dot{P} \sim$ 6$\times$10$^{-11}$ s/s (Woods et al. 1999b).

Woods et al (1999b) have suggested that the data may be consistent
with a discontinuous spin-down event during the second interval as a
result of the Superburst, as opposed to a doubling of $\dot{P}$ during
the entire second interval (Marsden, Rothschild and Lingenfelter 1999a).
This appears to be at odds with  the measurement of
$\dot{P}$=(11.0$\pm$1.7)$\times$10$^{-11}$ s/s in early June, 1998
(Kouveliotou et al. 1999), approximately 3 months before the Superburst
and with the RXTE/ASCA determination of $\dot{P} \sim$
10$\times$10$^{-11}$ s/s just after the event (Murakami et al. 1999).

\begin{figure}
\centerline{\psfig{file=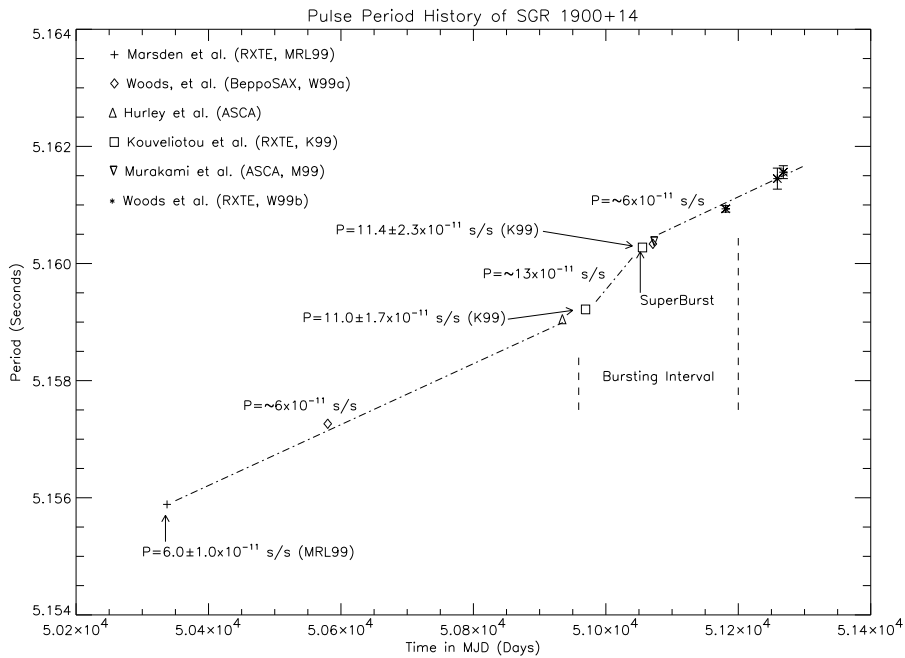, angle=90, width=12cm}}
\caption[]{The pulse period history for SGR 1900+14
versus time. All of the published values of the pulse period are given
along with the three measurements of $\dot{P}$ made as part of the
period determination analysis. The mean spin-down in the three time
intervals are also given.}
\end{figure}

\section{Spin-Down in SGR 1900+14 is Due to a Relativistic Wind}

Assuming that the spin-down torque is given by $\dot{\Omega} \propto
\Omega^n$,
the age of a pulsar with period $P$ and spin-down $\dot{P}$ is given by

\centerline{$ t_{age} = P/[(n-1)\dot{P}]$} 
\noindent where the spin-down braking index, $n$ = 3 for pure magnetic dipole
radiation and $n \sim$ 1 for wind torques.  Inverting the age equation
yields
 
\centerline{ $n = 1 + (P/\dot{P}) (t_{age})^{-1}$.} 
\noindent Using parameters appropriate for SGR 1900+14, we find that 

\centerline{ $n = 1 + 0.27/(t_{age}/10^4$yr).} 
\noindent This indicates that the braking index for SGR 1900+14 must be $\sim$1,
and that the spin-down of SGR 1900+14 is dominated by torques due to the relativistic wind and not magnetic dipole radiation.

\section{Spin-Down Torques of SGRs}

In reality, there will be more than one torque spinning down the pulsar
at any given time.
The torque provided by the emission of a relativistic wind is (Thompson et al. 1999):

\centerline{ $I_* \dot{\Omega}_w = -\Lambda(L_w/c^2)R_A ^2 \Omega$}
\noindent where $I_*$ is the neutron star moment of inertia, $L_w$ is
the luminosity of the wind, $\Omega \equiv 2\pi/P$ is the spin frequency,
$\dot{\Omega}_w$ is the spin-down rate due to the wind, and $R_A$ is the
Alfven radius.  $\Lambda$ is a constant equal to 2/3 for a magnetic
dipole field aligned with the rotation axis.
The Alfven radius is given by:

\begin{figure}[t]
\centerline{\psfig{file=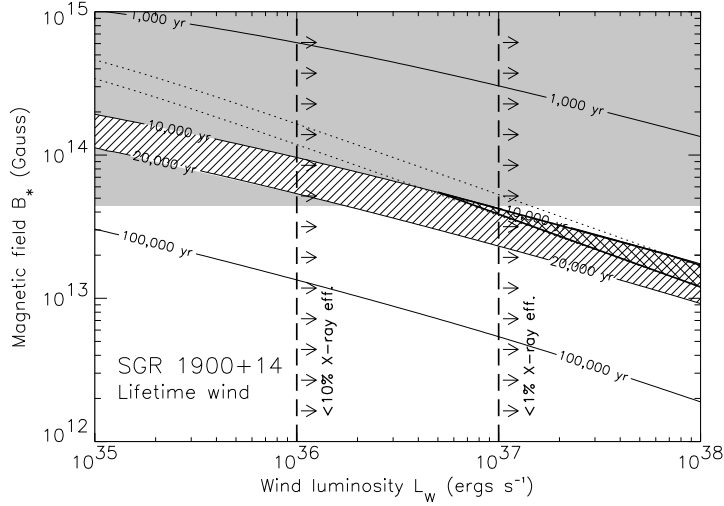, angle=0, width=10cm}}
\caption[]{Age contours for SGRs 1900+14 for constant magnetic field
and luminosity.  The cross-hatched areas denote the allowed regions of
parameter space given the constraints provided by the age of the
associated supernova remnant (solid lines) and long term present-day
spin-down rate (dotted lines). The vertical dashed lines denote the
10\% and 1\% efficiencies of the wind in producing the observed x-ray
flux of 10$^{35}$ ergs/s.}

\end{figure}

\centerline{ $\frac{L_w}{4 \pi R_A ^2 c} = \frac{B_*^2 (R_A)}{8 \pi}$}
\noindent where $B_*$ is the magnetic field of the neutron star.
When the Alfven radius is inside the light cylinder radius ($R_A < R_{lc}$, where $R_{lc} = c/\Omega$),

\centerline{ $I_* \dot{\Omega}_w = - \Lambda B_* R_*^3 \left( \frac{L_w}{2c^3} \right)^{1/2} \Omega$}
\noindent where $R_*$ is the radius of the neutron star and dipole geometry is assumed.
When the Alfven is outside the light cylinder radius, the torque is limited to

\centerline{$I_* \dot{\Omega}_w = - \Lambda L_w \Omega^{-1}$}
\noindent The transition frequency between these two wind spin-down regimes is 

\centerline{{ $\Omega_{tr} = 8.572 \left( \frac{L_w}{10^{36} \rm{ergs/s}} \right)^{1/4} \left( \frac{B_*}{10^{14} \rm{G}} \right)^{-1/2} $} radians/s.}
\noindent The torque due to a rotating magnetic dipole is (Shapiro \& Teukolsky 1983):

\centerline{ $I_* \dot{\Omega}_{mdr} = - k \frac{B_*^2 R_*^6}{6c^3} \Omega^3$}
\noindent where k = 1 (Harding, Contopoulos, \& Kazanas 1999).

Once the total spin-down torque is specified as a function of $\Omega$,
the age of the SGR can found by the integral of $d\Omega$ over the total
torque divided by $I_*$, where the integration is performed from an initial frequency to the present-day angular frequency.

\section{Magnetic Field and Wind Luminosity Limits}

Using the above model we explore a wide range of magnetic fields $B_*$
and wind luminosity $L_w$, shown in Fig. 2. We see that the presently
observed period of $P$ = 5.157 s, the spindown rate of $\dot{P} =
6\pm1\times10^{-11}$ s/s (dotted lines) of SGR 1900+14, and the 10 to
20 Kyr range of ages (solid lines) of its associated supernova remnant
G42.8+0.6 (Vaisht et al. 1994), tightly constrain the allowable
magnetic field to $B_* < 6\times10^{13}$ G and wind luminosities $L_w >
5\times10^{36}$ erg/s. Compared to the quiescent 2-10 keV x-ray
luminosity of $\sim 10^{35}$ erg/s (Murakami et al. 1999), this wind
luminosity implies a $<$ 2 \% conversion efficiency of wind energy to
x-rays in that band which is quite consistent with theoretical
calculations (Tavani 1994, Harding 1995; Harding, Contopoulos \&
Kazanas 1999).  The magnetic field limits are also quite consistent
with the limiting values inferred for radio pulsars, but not with those
expected for magnetars. Very similar limits are set by comparable
analyses (Harding, Contopoulos, \& Kazanas 1999; Marsden, Rothschild \&
Lingenfelter 1999b) of SGR 1806-20 and its supernova remnant
G10.0-0.3.

\end{document}